\documentclass[fleqn,usenatbib]{mnras}
\usepackage{subfig}
\usepackage{graphicx}
\usepackage{newtxtext,newtxmath}

\title[The multi-phase outflow of J1509+0434]{A near-infrared study of the multi-phase outflow in the type-2 quasar J1509+0434}
\author[C. Ramos Almeida et al.]
{\parbox{\textwidth}{C. Ramos Almeida$^{1,2}$\thanks{Ram\'on y Cajal Fellow. E-mail: cra@iac.es},
J. A. Acosta-Pulido$^{1,2}$, 
C. N. Tadhunter$^{3}$,
C. Gonz\'alez-Fern\'andez$^{4}$,
C. Cicone$^{5}$,
and M. Fern\'andez-Torreiro$^2$
}\vspace{0.4cm}\\
\parbox{\textwidth}{$^{1}$Instituto de Astrof\' isica de Canarias, Calle V\' ia L\'actea, s/n, E-38205, La Laguna, Tenerife, Spain\\
$^{2}$Departamento de Astrof\' isica, Universidad de La Laguna, E-38206, La Laguna, Tenerife, Spain\\
$^{3}$Department of Physics \& Astronomy, University of Sheffield, Sheffield S3 7RH, UK\\
$^{4}$Institute of Astronomy, Madingley Road, Cambridge CB3 0HA, UK\\
$^{5}$INAF - Osservatorio Astronomico di Brera, Via Brera 28, I-20121 Milano, Italy\\ 
}}

\begin{document}

\date{}

\pagerange{\pageref{firstpage}--\pageref{lastpage}} \pubyear{2019}

\label{firstpage}
\maketitle

\begin{abstract}
Based on new near-infrared spectroscopic data from the instrument EMIR on the 10.4 m Gran Telescopio Canarias (GTC) we report the presence of an ionized and warm molecular outflow in the luminous type-2 quasar J150904.22+043441.8 (z = 0.1118). The ionized outflow is faster than its molecular counterpart, although the outflow sizes that we derive for them are consistent within the errors (1.34$\pm$0.18 kpc and 1.46$\pm$0.20 kpc respectively). We use these radii, the broad emission-line luminosities and in the case of the ionized outflow, the density calculated from the trans-auroral [OII] and [SII] lines, to derive mass outflow rates and kinetic coupling efficiencies. Whilst the ionized and warm molecular outflows represent a small fraction of the AGN power ($\leq$0.033\% and 0.0001\% of L$_{bol}$ respectively), the total molecular outflow, whose mass is estimated from an assumed warm-to-cold gas mass ratio of 6$\times10^{-5}$, has a kinetic coupling efficiency of $\sim$1.7\%L$_{bol}$. Despite the large uncertainty, this molecular outflow represents a significant fraction of L$_{bol}$ and it could potentially have a significant impact on the host galaxy. In addition, the quasar spectrum reveals bright and patchy narrow Pa$\alpha$ emission extending out to 4\arcsec~(8 kpc) South-East and North-West from the active nucleus.
\end{abstract}





\section{Introduction}
\label{intro}
 
Active galactic nuclei (AGN) can affect the interstellar medium of their host galaxies by consuming, heating, sweeping out and/or disrupting the gas available to form new stars \citep{2012ARA&A..50..455F,2015Natur.521..192P}. Indeed, semi-analytic models and simulations of galaxy formation require this feedback from the AGN for quenching star formation therefore producing realistic numbers of massive galaxies \citep{2005Natur.435..629S,2006MNRAS.365...11C}. However, many observations reveal quasar-driven outflows with radial sizes $\leq$1--3 kpc when seeing-smearing effects are accounted for \citep{2016A&A...594A..44H,2018ApJ...856..102F,2018MNRAS.478.1558T} and the outflow kinetic powers measured for AGN of different luminosities vary by four orders of magnitude ($\sim$0.001--10\% of L$_{bol}$; \citealt{2018NatAs...2..198H}). The impact of the observed outflows on their host galaxies is thus far from having been constrained.


The problem is that the contribution from the different gas phases entrained in the winds have not been determined in unbiased and representative AGN samples \citep{2018NatAs...2..176C}. This is primarily due to the reduced wavelength coverage of the observations, generally restricted to the optical. Other important sources of uncertainty are the assumptions needed to derive outflow properties such as mass rate and kinetic power \citep{2018NatAs...2..198H}, generally used to compare with models and simulations.

Type-2 quasars (QSO2s; L$_{[OIII]}>$10$^{8.5}L_{\sun}$; \citealt{2008AJ....136.2373R}) are excellent laboratories to search for outflows and study their influence in their host galaxies. This is because the emission lines produced in the broad line region (BLR) and the AGN continuum are obscured by nuclear dust, making it easier to detect broad lines associated with the outflows and study the stellar populations of the host galaxies. In \citet{2017MNRAS.470..964R} we demonstrated the feasibility of near-infrared (NIR) spectroscopy to characterize QSO2 outflows in the ionized and warm molecular phases of the gas. We found that the outflow properties were different from those derived using optical data and gas phase dependent, as previously claimed by \citet{2013ApJ...775L..15R}. This implies that all single-phase estimates of outflow properties provide an incomplete view of AGN feedback \citep{2017A&A...601A.143F,2018NatAs...2..176C}. The NIR range does not only include emission lines tracing ionized and warm molecular gas, but it is less affected by extinction (A$_K\approx 0.1\times A_V$) and permits to reach lower seeing values than the optical. In spite of this, the NIR spectrum of nearby QSO2s remains practically unexplored \citep{2013ApJ...775L..15R,2015MNRAS.454..439V,2017MNRAS.470..964R}.


\begin{figure}
\includegraphics[width=9cm]{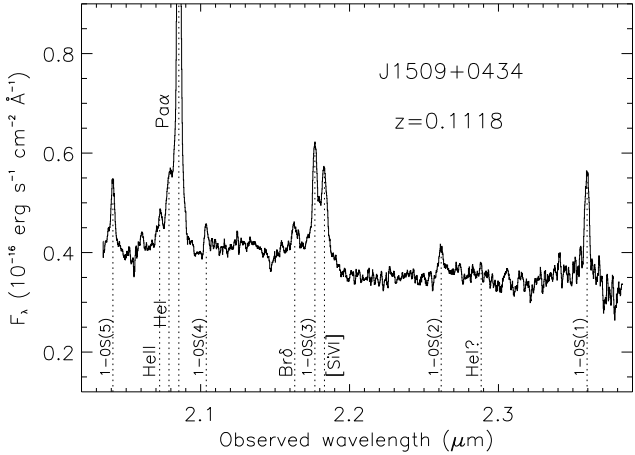}
\caption{Spectrum of the central region of J1509, extracted in an aperture of 0.8\arcsec~($\sim$1.6 kpc) and smoothed using a 6 pixel boxcar.}  
\label{fig1}
\end{figure}

We report high spectral resolution (R$\sim$4000) K-band spectroscopy of the QSO2 SDSS J150904.22+043441.8 (also IRAS F15065+0446 and hereafter J1509) obtained with the instrument Espectr\'ografo Multiobjeto Infra-Rojo (EMIR) on the 10.4 m Gran Telescopio Canarias (GTC). This is a nearby QSO2 (z = 0.1118) from the catalogue of \citet{2008AJ....136.2373R}, having an [O III] luminosity of 10$^{8.56}L_{\sun}$     (L$_{bol}$=4.9$\times10^{45}$ erg~s$^{-1}$ using the bolometric correction of \citealt{2004ApJ...613..109H}). It is a luminous IR galaxy (LIRG) according to its total IR luminosity log(L$_{IR}/L_{\sun}$)=11.6 \citep{2011ApJ...730...19S} and the Sloan Digital Sky Survey (SDSS) broad-band optical images show a disrupted galaxy morphology that could be indicative of a past galaxy interaction/merger.

Throughout this letter we assume a cosmology with H$_0$=71 km~s$^{-1}$ Mpc$^{-1}$, $\Omega_m$=0.27, and $\Omega_{\Lambda}$=0.73. At the distance of the galaxy (D$_L$=515 Mpc) the spatial scale is 2.017 kpc arcsec$^{-1}$.

\section{GTC/EMIR observations}

J1509 was observed with the NIR multi-slit spectrograph EMIR \citep{2006SPIE.6269E..18G,2014SPIE.9147E..0UG}, installed at the Naysmith-A focal station of the 10.4 m GTC at the Roque de los Muchachos Observatory, in La Palma. EMIR is equipped with a 2048$\times$2048 Teledyne HAWAII-2 HgCdTe NIR-optimized chip with a pixel size of 0.2\arcsec. 
We obtained a K-band (2.03--2.37 \micron) spectrum during the night of 2018 June 28th in service mode (Proposal GTC77-18A; PI: Ramos Almeida). The airmass during the observation was 1.26--1.38, the observing conditions were photometric and we estimated the seeing value by averaging the full-width at half maximum (FWHM) of eleven stars in the combined J-band acquisition image (0.76$\pm$0.07\arcsec). This is consistent with the FWHM=0.8\arcsec~measured from the K-band spectrum of the A0 star HD\,142346, observed after the science target to allow flux calibration and telluric correction. 
The slit width used during the observations was 0.8\arcsec, allowing a spectral resolution of $\sim$85 km~s$^{-1}$ at 2.1 \micron. The instrumental width measured from the OH sky lines is 5.8$\pm$0.2 \AA~with a dispersion of 1.71 \AA~pixel$^{-1}$. Slit losses were minimal ($\sim$5\% as estimated from the A0 star spectrum) thanks to the good seeing and photometric conditions during the observation.
J1509 was observed for a total on-source integration time of 1920 s following a nodding pattern ABBA. 
The two nodding positions were separated by 30\arcsec~and the slit was oriented along PA=-16$\degr$, centred on the galaxy nucleus and following the extended emission observed in the color-combined optical SDSS image of the galaxy.  
The data were reduced using the {\it lirisdr} software within the IRAF enviroment. Consecutive pairs of AB two-dimensional spectra were subtracted to remove the sky background. Resulting frames were then wavelength-calibrated and flat-fielded before registering and co-adding all frames to provide the final spectra. 
The wavelength calibration was done using the HgAr, Ne and Xe lamps available. 
From the sky spectrum we measured a wavelength calibration error of 8.33 km s$^{-1}$.



\section{Results}

\subsection{Nuclear spectrum}
\label{nuclear}

\begin{figure*}
\centering
{\par\includegraphics[width=5.6cm,angle=90]{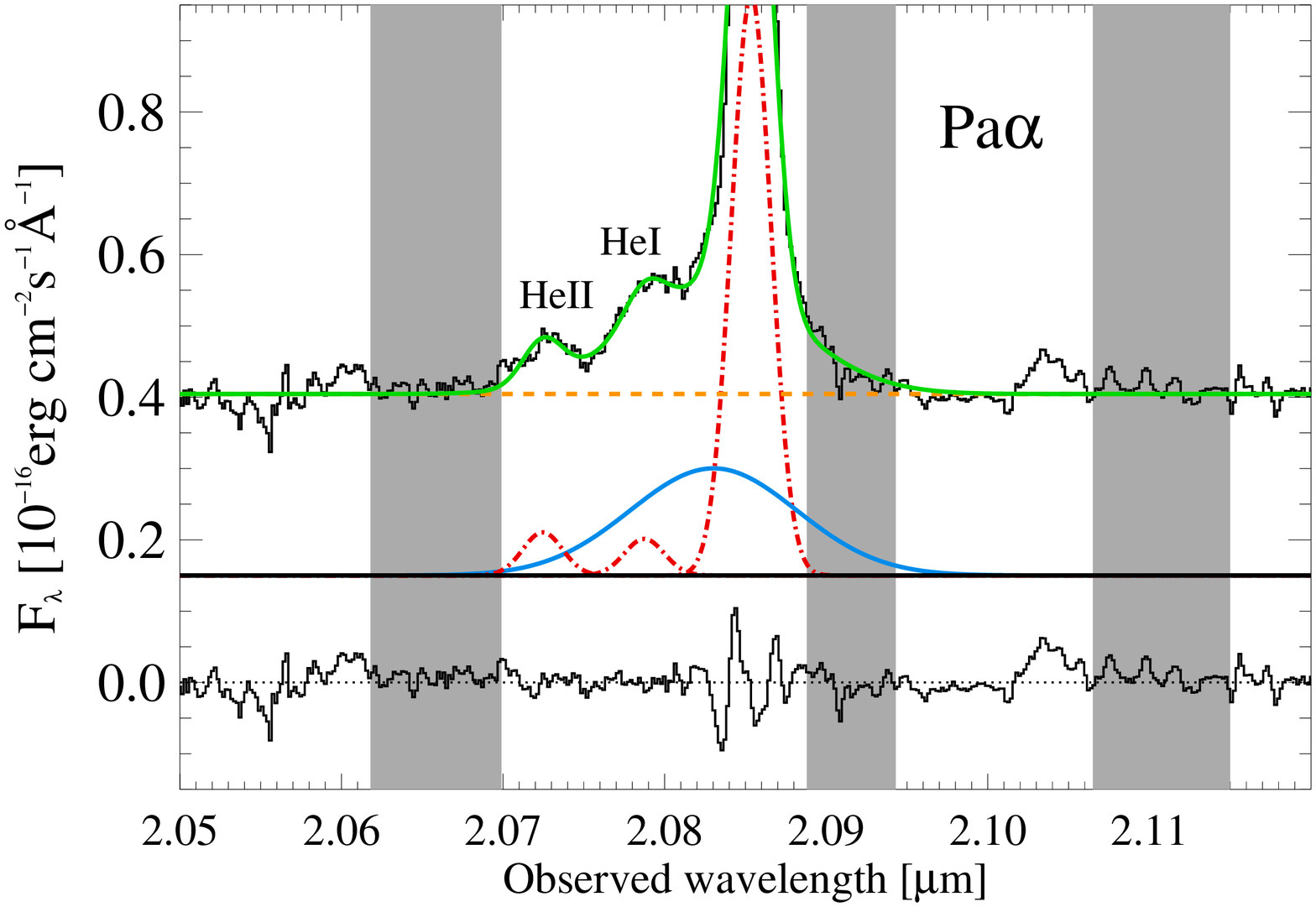}
\includegraphics[width=5.6cm,angle=90]{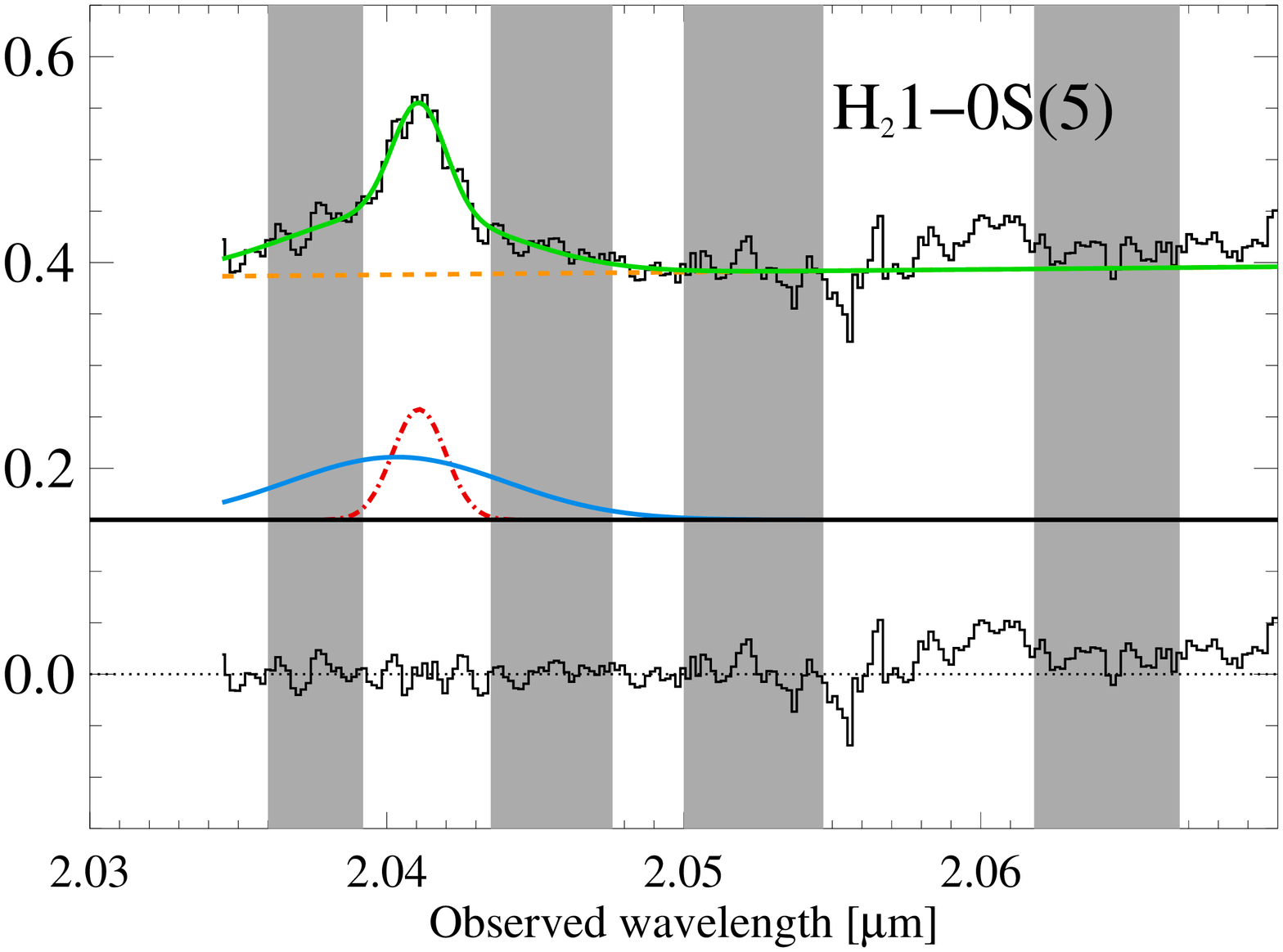}\par}
\caption{Examples of line profiles showing a blueshifted broad component. Solid blue, dot-dashed red and dashed orange lines correspond to broad, narrow components and continuum. Solid green lines are the total fits. The Gaussian components have been vertically shifted and the residuals of the fits are shown at the bottom of each panel. The shaded areas indicate the wavelength ranges employed in the determination of the outflow sizes.}
\label{fig2}
\end{figure*}

In order to study the nuclear emission of J1509 we extracted a spectrum in an aperture of 0.8\arcsec~(1.61 kpc), centred at the peak of the continuum emission. It was then flux-calibrated and corrected from atmospheric transmission by using the A0V star spectrum and the IRAF task {\it telluric}.  
The spectrum reveals several emission lines tracing different phases of the gas (see Fig. \ref{fig1}). We detect the low-ionization lines Pa$\alpha$, Br$\delta$, HeII$\lambda$1.8637\micron, HeI$\lambda$1.8691\micron~and tentatively HeI$\lambda$2.0587\micron. We also detect the high-ionization/coronal line [SiIV]$\lambda$1.9634\micron~(ionization potential = 166.7 eV) and all the molecular lines from H$_2$ 1-0S(5) to 1-0S(1). These H$_2$ lines trace warm molecular gas (T$>$1000 K), which is only a small fraction of the total molecular gas content \citep{2005AJ....129.2197D,2017A&A...607A.116E}. The H$_2$1-0S(3) line is partially blended with [SiVI]. 

The emission lines were fitted with Gaussian profiles using the Starlink package {\it DIPSO}. We used the minimum number of Gaussians necessary to correctly reproduce the line profiles leaving a flat residual (see Fig. \ref{fig2}). 
The ionized lines show a narrow component of $\sim$400--450 km~s$^{-1}$, typical of the narrow-line region (NLR), whilst those of the molecular lines are slightly narrower (300-350  km~s$^{-1}$).
Resulting FWHMs, velocity shifts (V$_s$) and fluxes are reported in Table \ref{tab1}. The FWHMs are corrected from instrumental broadening and V$_s$ are relative to the central wavelength of the Pa$\alpha$ narrow component ($\lambda_c$ = 20853.52$\pm$0.64 \AA), from which we determine a redshift of z = 0.11182$\pm$0.00003. Uncertainties in V$_s$ include the wavelength calibration error and individual fit uncertainties provided by {\it DIPSO}. Flux errors were determined by adding quadratically the flux calibration error ($\sim$5\% estimated from the A0 star) and the fit uncertainties.

In the case of Pa$\alpha$, Br$\delta$, [SiVI] and all the five H$_2$ lines detected in our nuclear spectrum, two Gaussians are needed to reproduce their asymmetric line profiles. One corresponds to the narrow component and the other to a blueshifted broad component that might be the approaching side of a biconical
outflow (see e.g. \citealt{2010ApJ...708..419C,2016ApJ...828...97B}). The latter are indicated with a (b) in Table \ref{tab1} and as blue solid lines in Fig. \ref{fig2}. The FWHM of the blueshifted components measured for Pa$\alpha$ is 1750$\pm$175 km~s$^{-1}$, with V$_s$=-335$\pm$70 km~s$^{-1}$. The same components are needed to reproduce the Br$\delta$ profile, of much lower intensity and partially blended with the blue wing of H$_2$1-0S(3). For this reason some of the input parameters were fixed to obtain a reliable fit (see Table \ref{tab1}).

For comparison with the NIR ionized lines we fitted the H$\beta$ and [OIII]$\lambda$5007 \AA~lines detected in the optical spectrum of J1509 publicly available from the SDSS data release 14 \citep{2018ApJS..235...42A}. 
We measured a FWHM$\approx$400 km~s$^{-1}$ for the narrow components of H$\beta$ and [O III].
The broad H$\beta$ component has a FWHM=1200$\pm$100 km~s$^{-1}$ and it is blueshifted by 445$\pm$80 km~s$^{-1}$ from the central wavelength of the narrow component. For the [OIII] line we fitted a broad component of FWHM=1500$\pm$20 km~s$^{-1}$, blueshifted by 350$\pm$10 km~s$^{-1}$. Thus, despite the different scales probed by the optical and NIR data (the SDSS spectrum corresponds to an aperture of 3\arcsec~in diameter) we find consistent results. 
The [SiVI] line detected in our NIR spectrum also shows a blueshifted broad component of FWHM=1450$\pm$300 km~s$^{-1}$ which indicates that the highly-ionized gas is also outflowing. This broad component is less blueshifted (-100 km~s$^{-1}$), although with relatively large uncertainty due to the blend with the H$_2$ line, than the low-ionization broad components, as we also found for the Teacup galaxy \citep{2017MNRAS.470..964R}. In the optical spectrum we find a blueshifted broad component of FWHM=1270$\pm$200 km~s$^{-1}$ for the coronal line [FeVII]$\lambda$6087 \AA~(IP=99.1 eV), consistent within the errors with the FWHM of [SiVI].
Finally, we detect broad components in the warm molecular lines. These broad components have FWHMs$\sim$1300 km~s$^{-1}$ and are blueshifted by $\sim$100 km~s$^{-1}$ relative to narrow Pa$\alpha$. For the majority of the H$_2$ lines we had to fix the input parameters of the broad component to obtain good fits (see Table \ref{tab1}), but these blueshifted broad components are necessary to reproduce the line profiles and produce flat residuals (see the right panel of Fig. \ref{fig2}). A significant fraction of the warm molecular gas is also outflowing.

\begin{figure*}
\centering
{\par\includegraphics[width=5.6cm,angle=90]{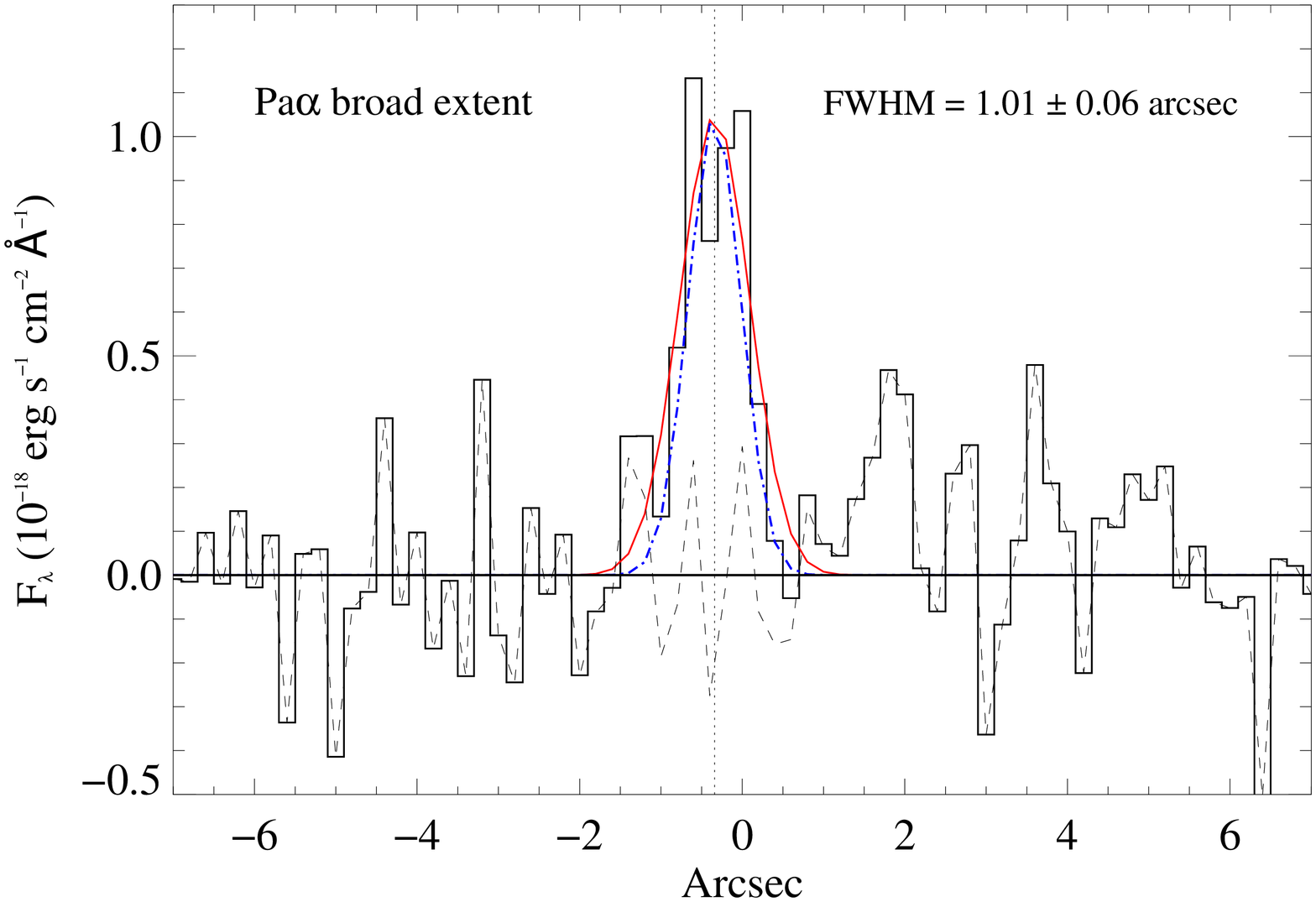}
\includegraphics[width=5.6cm,angle=90]{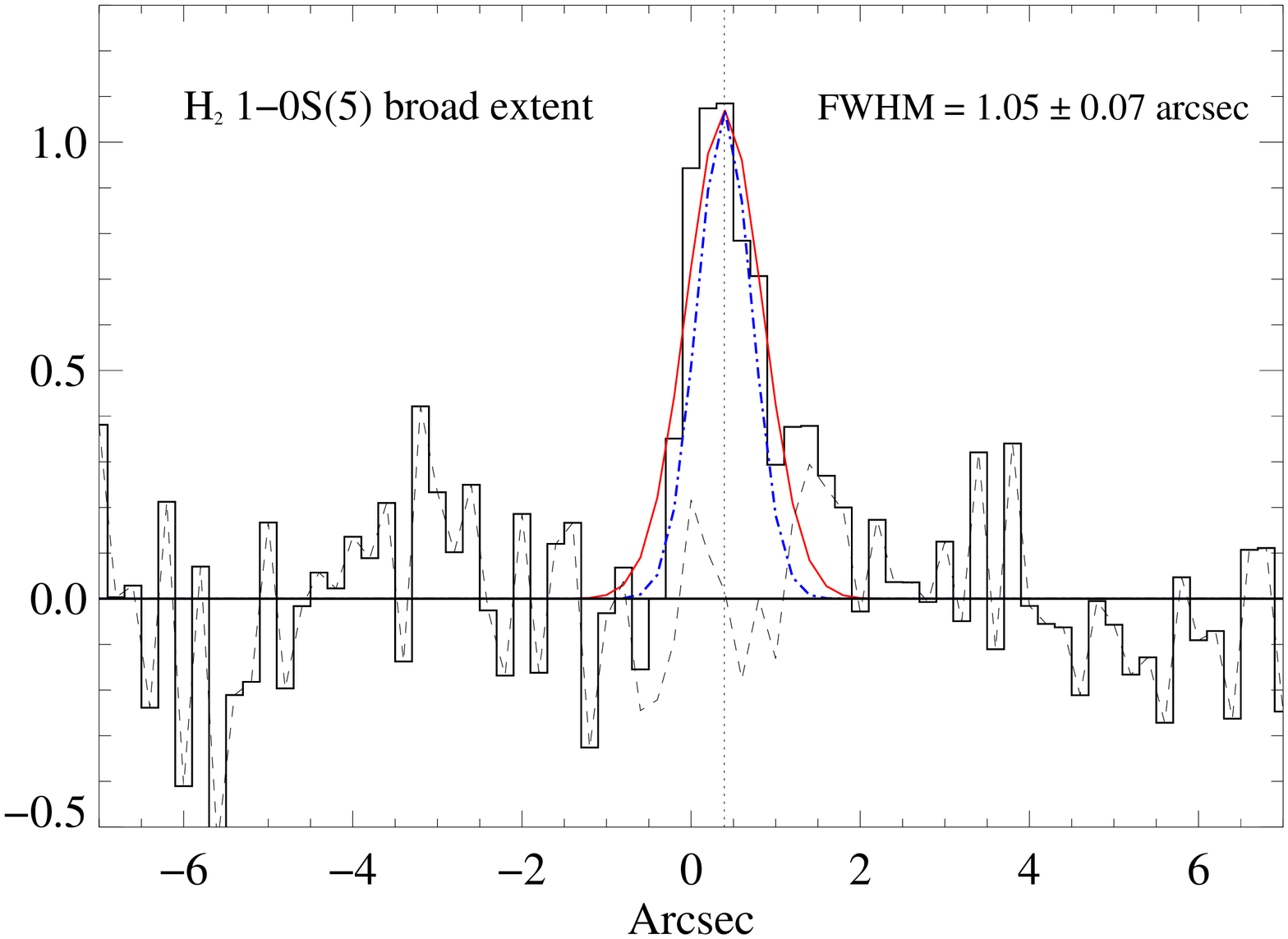}\par}
\caption{Continuum-subtracted spatial profiles of the broad components of Pa$\alpha$ and H$_2$1-0S(5) used to calculate the outflow sizes. The Pa$\alpha$ profile was obtained from the red wing of the line only, and the H$_2$ profile is the average of the profiles obtained from the blue and red wings. The red solid lines correspond to the fitted Gaussians and the blue dot-dashed line to the seeing FWHM.}
\label{fig3}
\end{figure*}

In order to constrain the spatial extent of the outflows we followed the methodology employed in \citet{2018MNRAS.474..128R}. We averaged spatial slices of the blue and red wings of the broad lines detected in the nuclear spectrum, avoiding the wavelength range covered by the corresponding narrow emission line and any other adjacent emission lines (e.g. the low-intensity helium lines blended with the blue wing of Pa$\alpha$; see Fig. \ref{fig2}). We did the same for spatial slices blueward and redward of the broad emission line, averaged them and subtracted from the broad line emission. By doing this we derive continuum-free spatial profiles of the gas in the outflow, so we can fit them with a Gaussian to measure FWHM$_{obs}$ (see Fig. \ref{fig3}). We consider the outflow resolved if
\begin{equation}
FWHM_{obs} > FWHM_{seeing} + 3\sigma = 0.76\arcsec + 3\times0.07 = 0.97\arcsec
\end{equation}
Finally we subtracted the seeing FWHM in quadrature to derive the outflow size
\begin{equation}
FWHM_{out}=\sqrt{FWHM^2_{obs}-FWHM^2_{seeing}}
\end{equation}

Using the red wing of the Pa$\alpha$ line we find that the ionized outflow in J1509 is barely resolved, and we measure FWHM$_{out}$=1.34$\pm$0.18 kpc. For the molecular outflow we used the S(5) and S(1) lines, which have the highest S/N in our nuclear spectrum. In the case of S(1) the noisy continuum prevents a good determination of the continuum-free spatial profiles, but using the continuum-subtracted red and blue wings of the S(5) line (only red continuum available; see Fig. \ref{fig2}) we also find the molecular outflow to be barely resolved, with FWHM$_{out}$=1.46$\pm$0.20 kpc. The errors were estimated by adding in quadrature the seeing error and the standard deviation of FWHM$_{obs}$ obtained from varying the wavelength range covered by the continuum and red and blue wings of the lines. 

\subsection{Extended emission}
\label{extended}

\begin{figure}
\includegraphics[width=8cm]{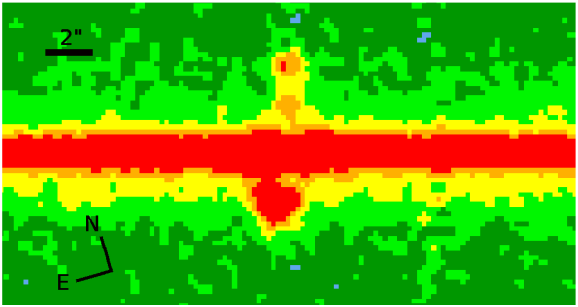}
\caption{Two-dimensional spectrum of J1509 centred in the Pa$\alpha$ line, smoothed using a 2 pixel boxcar. A bright emission-line knot at 2.2\arcsec~(4.4 kpc) SE of the active nucleus and NW patchy emission are detected.}  
\label{fig4}
\end{figure}

The slit was oriented following the morphological structures visible in the optical SDSS image of J1509 (PA=-16\degr). Our J-band acquisition image shows extended emission roughly in the same direction. The two-dimensional K-band spectrum reveals the bright and patchy extended Pa$\alpha$ emission shown in Fig. \ref{fig4}. Towards the south-east (SE) we detect a bright and compact line-emitting blob peaking at 2.2\arcsec~(4.44 kpc) from the nucleus (as measured from the position of the maximum of the AGN continuum). The north-west (NW) extended Pa$\alpha$ emission is patchy and it shows a bright knot at 3.4\arcsec~(6.86 kpc) from the AGN nucleus (see Fig. \ref{fig4}). We will refer to these regions as the SE and NW knots. In order to constrain the total extent of the narrow Pa$\alpha$ emission we analyzed the line profiles detected in adjacent spectra extracted in apertures of 0.8\arcsec~at both sides of the nucleus (as measured from the maximum of Pa$\alpha$ emission). We detect narrow Pa$\alpha$ emission up to 4\arcsec~(8 kpc) SE and NW of the nucleus.
We extracted two additional spectra of the same aperture (0.8\arcsec) centred at the peak of each knot. 
The Pa$\alpha$ emission of the SE knot can be reproduced with a single Gaussian of FWHM$\sim$110 km~s$^{-1}$ and centred practically at the same wavelength of the nuclear Pa$\alpha$ narrow component. In the NW knot the line profile appears double-peaked due to the residuals of a sky line at 2.0857 $\micron$ but it can be fitted with a single Gaussian of FWHM$\sim$120 km~s$^{-1}$ and redshifted by $\sim$60 km~s$^{-1}$ relative to the nuclear Pa$\alpha$ narrow component (see Table \ref{tab1}). Thus, the extended gas kinematics provide no evidence for outflowing gas on this scales, despite the clear presence of warm ionized gas, 
although we cannot rule out the presence of low brightness outflow components \citep{2018MNRAS.478.2438S}.

\begin{table}
\centering
\small
\begin{tabular}{lccc}
\hline
\hline
 &\multicolumn{3}{c}{Nuclear spectrum} \\
Line & FWHM & V$_s$ & Line flux $\times 10^{-16}$ \\
& (km~s$^{-1}) $  & (km~s$^{-1}$) & (erg~cm$^{-2}$~s$^{-1}$)\\
\hline
He II     	         &       400    & 18$\pm$49  &  1.86$\pm$0.41   \\
He I     	         &       400    & 91$\pm$65  &  1.58$\pm$0.49   \\
Pa$\alpha$      	 &  404$\pm$13  &   0$\pm$9   & 25.08$\pm$1.53   \\
Pa$\alpha$ (b)        	 & 1748$\pm$175 &-335$\pm$67  & 19.45$\pm$2.12   \\
Br$\delta$ 	 	 &      400     &  83$\pm$45  &  1.64$\pm$0.26   \\
Br$\delta$ (b)	 	 & 1783$\pm$214 &     -330    &  4.11$\pm$0.75   \\
\hline
$[Si~VI]$ 		 &  448$\pm$76  &  27$\pm$22  &  3.81$\pm$1.23   \\
$[Si~VI]$ (b)		 & 1457$\pm$324 & -99$\pm$143 &  9.41$\pm$2.02   \\
\hline
H$_{2}$ 1-0S(5)	 	 &  291$\pm$48  &   1$\pm$18  &  2.40$\pm$0.57   \\
H$_{2}$ 1-0S(5) (b)	 & 1265$\pm$253 &-110$\pm$59  &  5.60$\pm$0.81   \\
H$_{2}$ 1-0S(4)	 	 &  327$\pm$52  &  28$\pm$24  &  1.40$\pm$0.23   \\
H$_{2}$ 1-0S(4) (b)	 &      1300    &     -100    &  0.14$\pm$0.24   \\
H$_{2}$ 1-0S(3)	 	 &  359$\pm$31  &  27$\pm$15  &  4.54$\pm$0.61   \\
H$_{2}$ 1-0S(3) (b)	 &      1300    &     -100    &  6.51$\pm$1.74   \\
H$_{2}$ 1-0S(2)		 &       300    &  43$\pm$36  &  1.22$\pm$0.30   \\
H$_{2}$ 1-0S(2) (b)	 &      1300    &      -100   &  1.67$\pm$0.81   \\
H$_{2}$ 1-0S(1)		 &  345$\pm$39  &  57$\pm$19  &  5.35$\pm$0.86   \\
H$_{2}$ 1-0S(1) (b)	 &      1300    & -42$\pm$116 &  6.31$\pm$1.49   \\
\hline	   					 			    															 &\multicolumn{3}{c}{SE knot} \\
Pa$\alpha$      	 &  112$\pm$9  &   7$\pm$8   & 6.59$\pm$0.35   \\
\hline	  																					      
 &\multicolumn{3}{c}{NW knot} \\
Pa$\alpha$      	 &  118$\pm$12  &  57$\pm$9   & 1.99$\pm$0.15   \\
\hline	  	
\end{tabular}	
\caption{Emission lines detected in the nuclear, SE and NW knots spectra of J1509. Velocity shifts (V$_s$) are relative to the central 
$\lambda$ of the narrow Pa$\alpha$ component. 
Measurements without errors correspond to fixed parameters.}
\label{tab1}
\end{table}

\section{Discussion and conclusions}

We detect blueshifted broad components in different emission lines of the nuclear spectrum of the QSO2 J1509 that we identify with the approaching side of a biconical multi-phase outflow. Integral field observations are required to confirm this geometry. The characteristics of this outflowing gas are different in the ionized and warm molecular phases, highlighting the importance of this kind of studies to evaluate the impact of AGN feedback. 
The ionized outflow is faster (V$_s$=-330 km~s$^{-1}$) than the warm molecular outflow (V$_s\approx$-100 km~s$^{-1}$). This is also the case for the obscured quasar F08572+3915:NW (L$_{bol}\sim$5.5$\times$10$^{45}$ erg~s$^{-1}$) studied by  \citet{2013ApJ...775L..15R} in the NIR using integral field spectroscopy and for which they reported blueshifted H$_2$ gas velocities of up to -1700 km~s$^{-1}$ in the inner 400 pc of the quasar. These results are in contrast with what we found for the Teacup  \citep{2017MNRAS.470..964R}, in which we detected the ionized outflow but not its molecular counterpart. 

In order to evaluate the power of the ionized and warm molecular outflows we need to estimate accurate mass outflow rates (\.M) and kinetic powers (\.E). To do so, we first require a good estimate of the outflow density. Taking advantage of the optical SDSS spectrum of J1509 we can measure the total fluxes of the [SII]$\lambda\lambda$6716,6731 and [OII]$\lambda\lambda$3726,3729 doublets as well as of the trans-auroral [OII]$\lambda\lambda$7319,7331 and [SII]$\lambda\lambda$4068,4076 lines. By doing so we can determine the electron densities (n$_e$) and reddening of the outflow region simultaneously, following the method described in \citet{2018MNRAS.474..128R}. The trans-auroral ratios F(3726+3729)/F(7319+7331) and F(4068+4076)/F(6717+6731) are sensitive to higher density gas than the classical [SII] and [OII] doublet ratios \citep{2011MNRAS.410.1527H} and therefore more suitable for estimating outflow densities. 

By comparing our measured [OII] and [SII] ratios 
(0.83$\pm$0.02 and -1.27$\pm$0.05) with a grid of photoionization models computed with CLOUDY (C13.04; \citealt{2013RMxAA..49..137F}), we obtain Log n$_e$ (cm$^{-3}$) = 3.25$\pm^{0.11}_{0.15}$ and E(B-V)=0.45$\pm$0.04. These values are consistent with those obtained by \citet{2018MNRAS.474..128R} and \citet{2018MNRAS.478.2438S} using the same methodology employed here for a sample of 17 ULIRGs with nuclear activity, redshifts 0.04<z<0.2 and bolometric luminosities 43.4$\leq$Log L$_{bol}\leq$46.3. It has been suggested that part of the outflow mass could be contained in lower density gas not traced by the trans-auroral lines \citep{2017ApJ...835..222S}, but high electron densities (Log n$_e\sim$4.5) are also reported by \citet{2019arXiv190311076B} for the outflows of nearby type-2 AGN using an independent method. We note that our value of n$_e$ is a lower limit on the outflow density because we used total fluxes instead of broad line fluxes which we could not fit for the trans-auroral lines, and the densities estimated from the two sets of broad [SII] and [OII] lines are always equal or higher than those obtained from the total line fluxes \citep{2018MNRAS.474..128R,2018MNRAS.478.2438S}.

To calculate the ionized outflow mass rate (\.M) we used equations 1, 2 and 3 in \citet{2018MNRAS.474..128R}, n$_e$, r$_{out}$, v$_{out}$ (defined as the difference between the peak velocities of the broad and narrow component of each line reported in Table \ref{tab1}) and the reddening-corrected broad Pa$\alpha$ flux. The latter was obtained using the E(B-V) calculated above for the outflow region and the \citet{2000ApJ...533..682C} reddening law (A$_K$=0.18$\pm$0.02 mag). We derive an outflow mass M $\leq9\times10^5~M_{\sun}$ and \.M $\leq0.46~M_{\sun}yr^{-1}$. The latter is at the lower end of the range of values reported by \citet{2018MNRAS.474..128R} and \citet{2018MNRAS.478.2438S} for nearby ULIRGs. We note that outflow mass and derived quantities are upper limits because we consider n$_e$ to represent a lower 
limit on the true electron density (M=9$\times$10$^5~M_{\sun}\times$~10$^{3.25}$/n$_e$). 


Using the H$_2$ 1-0S(3) and 1-0S(1) broad line fluxes and equations 2 and 3 in \citet{2008A&A...482..215M} we calculate an excitation temperature T$_{ex}\approx$ 2000 K for the molecular gas in the outflow assuming local thermal equilibrium (LTE). Under these conditions we can use equation 1 in \citet{2017A&A...607A.116E}, the extinction-corrected 1-0S(1) flux and the luminosity distance of the QSO2 to estimate the molecular outflow mass. We obtain M=(1.0$\pm$0.2)$\times10^4$~M$_{\sun}$, which is a factor 90 lower than the mass in the ionized outflow. This difference between phases is considerably lower than the factor of 1600 reported by \citet{2013ApJ...775L..15R} for the QSO2 F08572+3915:NW, for which they measured a molecular outflow mass of 5.2$\times10^4$~M$_{\sun}$ and \.M=0.13~M$_{\sun}yr^{-1}$. For J1509 we estimate \.M=0.001~M$_{\sun}yr^{-1}$ using the mass, v$_{out}$ and r$_{out}$ measured for the warm molecular outflow. We note that although J1509 and F08572+3915:NW have roughly the same bolometric luminosities, the latter is a ULIRG in an on-going merger system hosting 
one of the most extreme outflows detected in CO \citep{2014A&A...562A..21C}. 

As we mentioned in Section \ref{nuclear}, the warm H$_2$ component of the outflow is just a small fraction of the total molecular gas content. We can then use the warm-to-cold gas mass ratio of 6$\times10^{-5}$ measured for two nearby LIRGs with and without nuclear activity observed in the NIR and the sub-mm \citep{2014A&A...572A..40E,2016A&A...594A..81P} to infer the total molecular outflow gas mass in J1509. We obtain M$_{H_2}$=(1.7$\pm$0.4)$\times10^8$~M$_{\sun}$. This error does not include the large uncertainty in the assumed warm-to-cold gas ratio, which is yet scarcely measured in outflows. For comparison, \citet{2018A&A...616A.171P} 
reported ratios of (2.6$\pm$1.0)$\times$10$^{-5}$ within the outflows of three nearby ULIRGs. These ratios lie at the upper end of values reported by \citet{2005AJ....129.2197D} for starburst galaxies and buried AGN (10$^{-7}$--10$^{-5}$). Fortunately, on-going ALMA CO observations of J1509 and other nearby QSO2s will soon permit us to quantify the cold molecular gas within the outflow and measure corresponding warm-to-cold gas ratios.
If we calculate the outflow mass rate of J1509 using the total H$_2$ mass we obtain \.M$_{H_2}$=23~M$_{\sun}yr^{-1}$.

We note that all the previous estimates of M and \.M are quite conservative because we do not consider projection effects in the employed velocities (v$_{out}$). More realistic values can be obtained by using maximum outflow velocities (v$_{max}$=v$_{out}-FWHM/2$ = -1200 and -750 km~s$^{-1}$ for the ionized and molecular outflow) as in \citet{2017A&A...607A.116E}. In this case we obtain \.M$\leq1.66~M_{\sun}yr^{-1}$ for the ionized outflow and 0.01 and 176~M$_{\sun}yr^{-1}$ for the warm and total molecular outflow. Finally, we can calculate the kinetic power of the ionized and molecular outflows as
\begin{equation}
\dot{E}=\frac{\dot{M}}{2}(v_{max}^2+3\sigma^2) 
\end{equation}

\noindent with $\sigma$=FWHM/2.355. For the ionized outflow we measure \.E$\leq1.6\times10^{42}$ erg~s$^{-1}$. Dividing this value by L$_{bol}$ estimated in Section \ref{intro} from the [OIII] luminosity we obtain F$_{kin}\leq$0.033\%, which is the power of the outflow as a fraction of L$_{bol}$ or the kinetic coupling efficiency. For the molecular outflow we calculate a kinetic power of \.E$_{warm}$=4.9$\times10^{39}$ erg~s$^{-1}$, which represent 0.0001\%L$_{bol}$. In principle, these low values of F$_{kin}$ would indicate that neither of the two phases of the wind are very relevant in terms of energetics, even considering that only $\sim$20--30\% of the outflow energy is kinetic according to simulations \citep{2018MNRAS.478.3100R} and that a fraction of this energy is used to work against the gravitational potential. Taking this into account, only $\sim$0.5\%L$_{bol}$ would be transmitted to the ionized and molecular outflows that we are characterizing here \citep{2018NatAs...2..198H}.
However, if we estimate the total molecular mass in the outflow from the warm molecular mass and we use the same outflow radius and kinematics to work out the kinetic energy (\.E$_{H_2}$=8.2$\times10^{43}$ erg~s$^{-1}$) we find F$_{kin}\sim$1.7\% L$_{bol}$. Thus, the molecular outflow in J1509 represents a significant fraction of L$_{bol}$ and it could potentially have a significant impact on the host galaxy. Despite the large uncertainty in the assumed warm-to-cold gas ratio, our value of 1.7\%L$_{bol}$ is in agreement with the coupling efficiencies derived from CO-based measurements of AGN of similar bolometric luminosities as J1509 ($\sim$0.5--3\%; \citealt{2015A&A...583A..99F,2015A&A...580A...1M}). 
 
We have demonstrated the feasibility of GTC/EMIR spectroscopy for deriving accurate multi-phase outflow properties and evaluate their potential impact on the host galaxy. The next step is targeting representative quasar samples to investigate how different galaxy properties (e.g. radio jets, shocks, morphologies) might be influencing the characteristics of the outflows and ultimately, how these outflows (i.e. AGN feedback) are affecting the host galaxies.


Based on observations made with the Gran Telescopio CANARIAS (GTC), instaled in the Spanish Observatorio del Roque de los Muchachos of the Instituto de Astrof\' isica de Canarias, in the island of La Palma. CRA acknowledges the Ram\'on y Cajal Program through project RYC-2014-15779 and the Spanish Plan Nacional de Astronom\' ia y Astrof\' isica under grant AYA2016-76682-C3-2-P. CC acknowledges funding from the European Union's Horizon 2020 research and innovation programme under the Marie Sklodowska-Curie grant agreement No 664931.
CRA and CT thank F. Santoro for providing the CLOUDY grids necessary for calculating the electron density. We finally acknowledge the anonymous referee for useful comments and suggestions that enabled us to improve this work.

\bibliographystyle{mnras}
\bibliography{biblio} 

\label{lastpage}

\end{document}